\documentstyle[12pt]{article}
\newlength{\extraspace}
\newlength{\extraspaces}
\catcode`\@=11
\def\numberbysection{\@addtoreset{equation}{section}
\def\theequation{\arabic{section}.\arabic{equation}}}

\begin{document}
\addtolength{\baselineskip}{.7mm}
\thispagestyle{empty}
\begin{flushright}
TIT-HEP-389 \\
%MU-??? \\
ICRR-Report-415-98-11 \\
{\tt hep-ph/9804279} \\
April, 1998 
\end{flushright}
\vspace{2mm}
\begin{center}
{\large{\bf   The Sliding Singlet Mechanism 
              with Gauge Mediated Supersymmetry Breaking 
}}
\vspace{5pt}
%
%{\large{\bf   (tentative) }} 
%
\\[5mm]
{\sc Yuichi Chikira}\footnote{
\tt e-mail: ychikira@th.phys.titech.ac.jp}   \\[2mm]
{\it Department of Physics, Tokyo Institute of Technology \\
Oh-okayama, Meguro, Tokyo 152-0033, Japan}  \hspace{2mm} \\
 and  \hspace{2mm} \\
{\sc Naoyuki Haba}\footnote{
\tt e-mail: haba@phen.mie-u.ac.jp}   \\[2mm]
{\it Faculty of Engineering, Mie University, \\
Mie, 514-0008, Japan } \hspace{2mm} \\
 and  \hspace{2mm} \\
{\sc Yukihiro Mimura}\footnote{
\tt e-mail: mimura@icrr.u-tokyo.ac.jp}  \\[2mm]
{\it Institute of Cosmic Ray Research, University of Tokyo \\
Midori-Cho, Tanashi, Tokyo 188-8502, Japan} \\[5mm]
{\bf Abstract}\\[5mm]
{\parbox{14cm}{\hspace{5mm}
%
%  Abstract !!
%
%
The sliding singlet mechanism is 
one of the most interesting solutions of the 
triplet-doublet splitting problem. 
We analyze this mechanism in the 
gauge mediated supersymmetry breaking scenario.
We show that
the sliding singlet mechanism does not work 
in the naive gauge mediation scenario
 because of the singlet linear terms derived from 
gravity, 
although $F$ term is much 
smaller than the one in 
the gravity mediation scenario.
We also consider the extension in order for the sliding singlet 
mechanism to work.
 }}
\end{center}
\vfill
\newpage
\setcounter{section}{0}
\setcounter{equation}{0}
\setcounter{footnote}{0}
\def\theequation{\arabic{section}.\arabic{equation}}
%
%
%%%%%%%%%  Introduction %%%%%%%%%%%%%%%%%%%%%%%%%%%%%%%%%%%%%%
%
\vspace{7mm}
\pagebreak[3]
\addtocounter{section}{1}
\setcounter{equation}{0}
\setcounter{subsection}{0}
\setcounter{footnote}{0}
\begin{center}
{\large {\bf \thesection. Introduction}}
\end{center}
\nopagebreak
\medskip
\nopagebreak
\hspace{3mm}

Supersymmetric theories now stand as the most promising 
candidates for a theory beyond the standard model.
The minimal supersymmetric extension of the standard model 
naturally solves the gauge hierarchy problem and 
makes the three gauge couplings unify 
at the scale of $O(10^{16})$ GeV. 
Therefore, it suggests to us the idea of 
a grand unified theory (GUT). 
However, if we consider GUT, 
a new fine-tuning problem, the so-called 
triplet-doublet splitting problem, appears.
Therefore the colored triplet Higgs 
must be superheavy to avoid the rapid proton decay, 
while the doublet Higgs must have the mass 
of weak scale. 
Several ideas to solve this serious problem 
have been proposed. 
These are, for examples, 
the sliding singlet mechanism \cite{sliding, Neme}, 
the missing partner mechanism \cite{missing}, 
the Dimopoulos-Wilczek mechanism \cite{DW}, 
and the GIFT mechanism \cite{GIFT}. 
The sliding singlet mechanism is 
the simplest idea in which triplet-doublet splitting is
realized dynamically. 
When the singlet shifts to the potential minimum, the triplet-doublet 
splitting is realized automatically.
The linear term of the singlet can 
produce the suitable hierarchy between the weak and the 
GUT scale in the supersymmetric limit \cite{Neme}. 
Since the electro-weak symmetry breaking 
occurs at the tree level, 
this model is not the so-called radiative electro-weak
symmetry breaking scenario \cite{radiative}. 

\par
How does the situation change when supersymmetry breaking is
 switched on?
If the supersymmetry breaking occurs at high energy, such as 
in the gravity mediation scenario, 
the radiative corrections of K\"ahler potential
induce the doublet Higgs scalar mass, 
which is the so-called $B$ term of 
$O( \langle F \rangle M_{GUT}/M_P)$.
It destroys the Higgs mass hierarchy \cite{PS}.
One approach to avoid this difficulty is 
to extend the gauge symmetry from $SU(5)$ to 
$SU(6)$ \cite{Sen}.
Another approach is to consider the low energy 
supersymmetry breaking \cite{Neme, GM}.
The authors of Ref. \cite{DNS, CiPo} predicted that 
the sliding singlet mechanism may work 
in the gauge mediation scenario.
In this paper we analyze whether the sliding singlet mechanism 
can really work 
in the gauge mediation scenario or not. 

\par
Through the K\"ahler potential, 
the supersymmetry breaking effects induce 
the singlet linear terms both in the superpotential 
and in the soft supersymmetry breaking interactions \cite{Nilles}. 
We show that 
the sliding singlet mechanism does not work 
in the naive gauge mediation scenario 
because of these singlet linear terms, 
although the $F$ term is much 
smaller than the one in 
the gravity mediation scenario.
In order for the sliding singlet mechanism to work, 
additional extensions are needed. 
One of the extensions considered is 
the introduction of the additional 
strong gauge dynamics as will be shown.
Even if we introduce these extensions, 
we also need one more additional mechanism 
that induces the sliding singlet soft breaking mass of 
the order of soft breaking masses of Higgs for 
the electro-weak vacuum stability.

\par
This paper is organized as follows:
In section 2, we review 
the sliding singlet mechanism. 
Next, we estimate the linear terms of the singlet, 
which are induced by the gravitational interactions.
Section 3 is devoted to the analysis of 
the Higgs potential both at the GUT scale and 
at the messenger scale. 
In section 4 we give summary and discussions of these results.

%
%%%%%%%%%  Materials and Methods  %%%%%%%%%%%%%%%%%%%%%%%%%%%
%
\vspace{7mm}
\pagebreak[3]
\addtocounter{section}{1}
\setcounter{equation}{0}
\setcounter{subsection}{0}
\setcounter{footnote}{0}
\begin{center}
{\large {\bf \thesection. The Sliding Singlet Mechanism}}
\end{center}
\nopagebreak
\medskip
\nopagebreak
\hspace{3mm}

In this section, we review the sliding singlet mechanism
\cite{sliding, Neme}
and present our framework.
In GUT, we have to introduce 
colored Higgs triplets, $H_C$ and $\bar H_C$, to
embed the Higgs doublets in $SU(5)$ fundamental representations, 
$H = (H_C, H_u)$ and $\bar H = (\bar H_C, \varepsilon H_d^T)$.
However, the colored Higgs cannot be light 
because we do not want to have a proton decay that is too fast
or to spoil the successful unification of gauge couplings.
Thus, we need to split the Higgs doublets and triplets.
There are various attempts to solve the splitting problem
\cite{sliding, Neme, missing, DW, GIFT}.
The sliding singlet mechanism is one of these attempts.

One considers a superpotential for the Higgs fields and an adjoint
field $\Sigma$ of the following form, 
\begin{equation}
W = m_H \bar H H + \lambda' \bar H \Sigma H.
\end{equation}
The adjoint field $\Sigma$ breaks $SU(5)$ down to
$SU(3) \times SU(2) \times U(1)$.
So, we choose the desired vacuum state, 
\begin{equation}
\langle \Sigma \rangle = \sigma
\left(\begin{array}{ccccc}
      2 &&&& \\
      & 2 &&& \\
      && 2 && \\
      &&& -3 & \\
      &&&& -3 
      \end{array}
\right).
\end{equation}
Here $\sigma$ has a value of the order of GUT scale.
To split the doublets and triplets, we have to
tune the mass parameter to 
\begin{equation}
m_H = 3 \lambda' \sigma.
\end{equation}
This fine-tuning is an unattractive feature of the minimal 
$SU(5)$ GUT.
To avoid this fine-tuning, 
it was suggested that one replace the mass parameter by a singlet $S$ 
\cite{sliding}, 
\begin{equation}
 W = \lambda S \bar H H + \lambda' \bar H \Sigma H.
\label{thesuperpotential}
\end{equation}
The vacuum expectation value of the singlet will 
slide to a GUT scale, 
because of the $F$-flat conditions, 
\begin{equation}
    \frac{\partial W}{\partial H}= \bar H (\lambda S+ \lambda' \Sigma) 
    =0, 
\end{equation}
\begin{equation}
    \frac{\partial W}{\partial \bar H}= (\lambda S+ \lambda' \Sigma)
    H= 0.
\end{equation}

A question arises whether this device is stable \cite{Neme, PS}.
It is necessary that $H$ and $\bar H$ have non-zero vacuum expectation 
values at the GUT scale for successful sliding.
We know that the doublet Higgs will have vacuum expectation
values at the weak scale, 
for example, via a radiative electro-weak symmetry 
breaking scenario \cite{radiative}.
However, one cannot assure that the doublet Higgs 
have the vacuum expectation values of weak scale
in the context of the sliding singlet mechanism \cite{Neme}.
Nemeschansky \cite{Neme} discussed the models which have 
a linear term of the sliding singlet in 
the superpotential (\ref{thesuperpotential}), 
\begin{equation}
W = \lambda S \bar H H + \lambda' \bar H \Sigma H - L S.
\end{equation}
Then we 
obtain the supersymmetric vacuum as follows, 
\begin{equation}
\langle S \rangle = 3 \frac{\lambda'}{\lambda} \sigma, 
\end{equation}
\begin{equation}
\langle \bar H \rangle 
=  {}^T \langle  H \rangle  
= (0,0,0,0,\frac{v}{\sqrt2}),\quad v^2 = 2L/\lambda.
\label{Hvev}
\end{equation}
The colored Higgs mass is
\begin{equation}
M_{H_C} = 5 \lambda' \sigma, 
\end{equation}
and the doublet Higgs mass is just zero.

In general, however, quadratically divergent tadpole terms 
associated with singlets will arise 
in softly broken supersymmetric theory \cite{Nilles}, 
even if the K\"ahler potential is minimal.
The tadpole terms arise
due to 
supergravity corrections from operators
suppressed by the Planck mass.
The $\ell$-loop induced tadpole term 
is written in the following typical form, \cite{Nilles}
\begin{equation}
    {\cal L} \sim
\frac{N}{(16\pi^2)^\ell} \frac{\Lambda^2}{M_P} \int d^4 \theta
 e^{K/M_P^2} (S + S^\dagger),
\label{gravitational}
\end{equation}
where $N$ is the number of light chiral superfields
that appear in the loops and $\Lambda$ is the
cutoff for the quadratic divergence.
We make the reasonable assumption that $\Lambda \sim M_P$.
The sliding singlet communicates to the supersymmetry
breaking sector due to the superspace density, $e^{K/M_P^2}$
in Eq. (\ref{gravitational}).
The superspace density has the expansion, 
\begin{equation}
    e^{K/M_P^2} = 1 + \frac1{M_P^2}
    \left\{ \theta^2 K^i F_i + \bar \theta^2 K_{i*} F^{i*}
           + \theta^2 \bar \theta^2 
             (K_{ij*} + \frac{K_i K_{j*}}{M_P^2}) F^i F^{j*}
    \right\},
\end{equation}
where $K_i$ is the derivative of the K\"ahler potential
with respect to a supersymmetry breaking spurious superfield $Z^i$,
which has vacuum expectation values in the scalar and the $F$ component,
\begin{equation}
Z = \langle Z \rangle + \theta^2 \langle F_Z \rangle .
\end{equation}
We can omit the angles as long as there are no ambiguities.
The Lagrangian (\ref{gravitational}) is then
\begin{equation}
    {\cal L} \sim \frac{N}{(16\pi^2)^\ell} \left\{
        \int d^2 \theta \frac{K_{i*} F^{i*}}{M_P} S
        + (K_{ij*} + \frac{K_i K_{j*}}{M_P^2}) \frac{F^i F^{j*}}{M_P} S
        + h.c. \right\}.
\label{dangerous}
\end{equation}

The tadpole term seen in Eq. (\ref{dangerous}) can spoil 
the weak scale hierarchy \cite{PS, CiPo}.
In gravity mediated supersymmetry breaking, 
we should choose the $F$ component of the spurious fields
to be
$F_Z = M_W M_P$ 
to obtain the gravitino mass in the order of
weak scale $M_W$.
Therefore, the coefficient of the sliding singlet 
in the equation (\ref{dangerous})
is proportional to $F_Z^2 / M_P = M_W^2 M_P$.
This is much larger than the weak scale, thus
the sliding singlet mechanism 
is not stable in the gravity mediation model.
On the other hand, in the gauge mediation model
low energy supersymmetry breaking parameters are given by the
ratio $F_X/X$ ($\equiv \Lambda_{mes}$),
which is about $10^4$-$10^5$ GeV.
The chiral superfield $X$ is the spurious field
in the messenger sector.
Though the determination of $F_X$ in the messenger sector
originating from $F_Z$ depends on models,
$F_Z$ can be in general much smaller than the gravity mediation
model.
Thus the sliding singlet mechanism may work in gauge mediation
models.
It is worth studying the sliding singlet mechanism 
in the context of gauge mediation models in detail.

%On the other hand, in the gauge mediation model
%we can choose the value of $F_X$ to be much smaller 
%than gravity mediation model, 
%because the parameter of the messenger sector
%is given by only the ratio $Lambda_{\rm mes} = F_X/X$, which is $10^4-10^5$ 
%GeV.
%
%Thus, the sliding singlet mechanism may 
%work in the gauge mediation model \cite{Neme, CiPo}.
%It is worth studying the sliding singlet mechanism 
%in the context of gauge mediation model in detail.

In the gauge mediation model where the $F$ component
arises radiatively originating from the 
$F$ component ($F_Z$) in the supersymmetry breaking sector,
we find that $F_X = O(10^{-4} {\rm -} 10^{-5}) F_Z$
and that $X \sim \sqrt{F_X} \sim 10^4 {\rm -} 10^5$GeV,
namely, $F_Z = O(10^{12} {\rm -} 10^{15}) {\rm GeV}^2$.
Requiring the coefficient of the sliding singlet in equation
(2.14) to be less than the weak scale for the case of $\ell = 0$ ($\ell =2$),
we find $F_Z$ to be less than $10^{12}$ GeV$^2$
($10^{15}$ GeV$^2$).
In the next section, we find 
another constraint in aspect of 
the minimization of scalar potential.

%
%
%%%%%%%%%  Results  %%%%%%%%%%%%%%%%%%%%%%%%%%%%%%%%%%%%%%%%%
%
\vspace{7mm}
\pagebreak[3]
\addtocounter{section}{1}
\setcounter{equation}{0}
\setcounter{subsection}{0}
\setcounter{footnote}{0}
\begin{center}
{\large {\bf \thesection. Feasibility of the Sliding Singlet Mechanism
with Gauge Mediation
}}
\end{center}
\nopagebreak
\medskip
\nopagebreak
\hspace{3mm}

In this section, we study the
feasibility of implementing the gauge mediated supersymmetry
breaking scenario in the framework of the sliding singlet mechanism.
We perform the minimization analysis
above the messenger scale in the Wilsonian scheme.
The supersymmetry breaking parameters coming from the
messenger loops are highly suppressed, 
and are neglected in the analysis.
However, supergravity induced breaking parameters
will appear.
The scalar potential is then written as
\begin{eqnarray}
    V &\!\!\!=&\!\!\! \sum m_i^2 |\phi_i|^2    \nonumber \\
 &\!\!\!+& \!\!\!
     |\lambda \bar H H - L|^2 +
     |(\lambda S + \lambda' \Sigma) H |^2 +
     |\bar H (\lambda S + \lambda' \Sigma) |^2 \\
 &\!\!\!+& \!\!\!
     ( - \rho S + h.c.) + \mbox{$D$-terms} \nonumber \\
 &\!\!\!+& \!\!\!
     \mbox{$A$, $B$-terms}. \nonumber
\end{eqnarray}
In the scalar potential, the first term
represents the supergravity induced breaking mass term
for the scalar fields $\phi_i = \{S, H, \bar H\}$, 
which are relevant in this analysis.
The breaking masses $m_i^2$ are nearly equal to the 
gravitino mass.
%, which is equal to $F_X/\sqrt3 M_P$.
The final terms are scalar trilinear terms and bilinear terms.
%which for simplicity we ignore hereafter;
%they do not change our results substantially.
%
%
%
%Reading out from Eq. (\ref{dangerous}), 
%we find\footnote
%{The renormalization group flow will modify these
%relations between Eq. (\ref{Lsim}) and (\ref{rhosim}).
%We neglect the modification here.}
%
%\begin{equation}
%    L \sim - \frac{N}{(16 \pi^2)^\ell} \frac{M_P}{B} m_{3/2}^2, 
%\label{Lsim}
%\end{equation}
%and 
%\begin{equation}
%   \rho \sim \frac{N}{(16 \pi^2)^\ell} M_P m_{3/2}^2, 
%\label{rhosim}
%\end{equation}
%where we define the parameter $B$ as $B = F_X/X$.
The gravitino mass and the parameter $\rho$
is given by the full supersymmetry breaking 
order parameter $F_Z$ of the complete theory.
The parameter $\rho$ is given as
\begin{equation}
\rho \sim \frac{N}{(16\pi^2)^\ell} M_P m^2_{3/2}. 
\label{rhosim}
\end{equation}
The parameter $L$ includes an expectation value of spurious fields 
in the supersymmetry breaking sector, $Z$,
\begin{equation}
L \sim - \frac{N}{(16 \pi^2)^\ell} Z m_{3/2}. 
\label{Lsim}
\end{equation}
We define the ratio $B \equiv \rho/L$.
This parameterization is convenient especially in the direct gauge mediation
model, in which $B$ equals $\Lambda_{mes}$.

The linear term in the scalar potential
slides the vacuum expectation
values of the sliding singlet 
and induces the so-called $\mu$ term, which is a doublet Higgs mass 
parameter in the superpotential.
Denoting the vacuum expectation values of the Higgs fields 
in the same way as the Eqs. (\ref{Hvev}), 
we obtain the scalar potential as\footnote{We neglect 
$A,B$ terms because they do not
change our main result substantially.}
\begin{equation}
    V(S, v) =  | \lambda S - 3 \lambda' \sigma |^2 |v|^2
             + |\frac{\lambda v^2}2 -L |^2 + m_{3/2}^2 (|S|^2 + |v|^2)
             + (- \rho S + c.c.).
\label{scalarpot}
\end{equation}
The extremization conditions are
\begin{eqnarray}
\frac{\partial V}{\partial S}
&=&
 2 (\lambda S - 3 \lambda' \sigma) |v|^2 + 2 m_{3/2}^2 S - 2 \rho = 0, \\
\frac{\partial V}{\partial v}
&=&
 v \left( 2 |\lambda S - 3 \lambda' \sigma|^2 +  \lambda (\lambda |v|^2 - 2 L)
     + 2 m_{3/2}^2 \right) =0.
\label{parent}
\end{eqnarray}
Note that it is necessary 
for successful sliding that the Higgs fields have non-zero vacuum 
expectation values.
We obtain the $\mu$ parameter as
\begin{equation}
    \mu =  \lambda S - 3 \lambda' \sigma 
        =  \frac{\rho}{\lambda v^2} - 
           \frac{3 \lambda' m_{3/2}^2 \sigma}{\lambda^2 v^2}
 \quad (v \neq 0).
\label{find_mu}
\end{equation}
Substituting $\mu$ into equation (3.6), 
we obtain the following cubic equation with respect to $v^2$, 
\begin{equation}
    (\lambda^2 v^2)^3 - 2 \lambda L (\lambda^2 v^2)^2
  + 2 ( \lambda \rho - 3 \lambda' m_{3/2}^2 \sigma)^2 =0.
\label{cubic_eq}
\end{equation}
This equation yields a minimization solution for $v \neq 0$
only when the following condition is satisfied
\begin{equation}
 (\rho - 3 \frac{\lambda'}{\lambda} m_{3/2}^2 \sigma)^2 
< \frac{16\lambda}{27} L^3.
\label{rhoLcondition}
\end{equation}
The minimization solution lies in the range\footnote{
The extremization condition has another solution, 
but this solution is on the saddle point.}
\begin{equation}
    \frac{4L}{3 \lambda} < v^2 < \frac{2L} \lambda.
\end{equation}
Substituting Eqs. (\ref{Lsim}) and (\ref{rhosim}), 
we find 
\begin{equation}
    m_{3/2}^2 \stackrel{_{\textstyle >}}{_{\displaystyle \sim}}
    \frac{B^3}{M_P}, \quad v \stackrel{_{\textstyle >}}{_{\displaystyle \sim}} B.
\label{consequence}
\end{equation}
In models where $F_X$ arises at the messenger level directly
from a O'Raifeartaigh mechanism (a direct gauge mediation model),
the ratio $B$ is determined as  $F_X/X (= \Lambda_{mes})$,
and the vacuum expectation values of the Higgs field are
larger than 10 TeV.
%This means that the gravitino mass is
%larger than 1 MeV and that the vacuum 
%expectation value of Higgs field
%is larger than 10 TeV.
%
If the condition that the gravitino mass
is larger than 1 MeV is not satisfied,
 the sliding singlet mechanism does not work.
Further the doublet Higgs becomes of the order of the GUT scale mass
and are integrated out.

It is disastrous that
the vacuum expectation value of Higgs is
larger than 10 TeV.
One may consider that the vacuum expectation value
will be modified at lower energy.
However, the consequence will not change drastically.
The running of the parameter 
$L$ will be negligibly small\footnote{
The renormalization group equation with respect to $L$ is
\begin{equation}
    \frac{d L}{d (\log \mu_r)} = \frac{\lambda^2}{8 \pi^2} L,
\end{equation}
where $\mu_r$ is the renormalization scale.
}.
As another origin, a non-gravitational tadpole term arises 
with three-loops in which colored Higgs circulate.
We estimate the contribution as 
\begin{equation}
\frac{\alpha_3^2 \lambda}{(16 \pi^2)^2} 
\frac1{M_{H_c}} \int d^4 \theta
S X X^\dagger.
\end{equation}
However, it is unnatural that 
this operator just cancels out the parameter $L$ 
coming from the gravitational contribution.

Below the messenger scale, 
additional supersymmetry breaking terms will arise 
due to the messenger loops, 
but they are less than the order of 100 GeV and 
are negligible compared to 10 TeV.
The mass of the sliding singlet is of the order of 
the vacuum expectation value of the Higgs field.
Thus, the sliding singlet survives to lower energy.
After GUT particles decouple, 
the superpotential can be written as\footnote{
Note that we should consider the constant term, 
$\langle W \rangle \sim - L \langle S \rangle$, 
in the framework of supergravity.
One need another sector
to wipe out the cosmological constant.
}
\begin{equation}
W =  (\mu + \lambda \tilde S) H_d H_u - L \tilde S,
\end{equation}
where $\tilde S = S - \langle S \rangle$.
The supersymmetry breaking terms are 
\begin{equation}
V_{soft} = m^2_{H_d} | H_d |^2 + m^2_{H_u} | H_u |^2
+ m^2_S | \tilde S |^2 
+ (\lambda A_\lambda \tilde S H_d H_u 
   + B_\mu \mu H_d H_u - \rho \tilde S 
  + h.c.).
\end{equation}

We denote the vacuum expectation values of the Higgs doublets
by $v_d$ and $v_u$, and the value of the singlet $\tilde S$ by $x$.
As a function of these vacuum expectation values, the scalar potential has the form
\begin{eqnarray}
V &=& | \lambda v_d v_u - L|^2 
+ |\mu + \lambda x|^2 (|v_d|^2 + |v_u|^2 ) \nonumber \\
&&
+ m^2_{H_d} |v_d|^2 + m^2_{H_u} |v_u|^2 + m_S^2 |x|^2  \\
&& + (\lambda A_\lambda v_d v_u x + B_\mu \mu v_d v_u - \rho x + h.c.)
  + \frac{\bar g^2}8 ( |v_d|^2 - |v_u|^2)^2, \nonumber
\end{eqnarray}
where $\bar g^2 = g_2^2 + {g'}^2$.

The extremization conditions can be written as 
\begin{equation}
{\mu'}^2  + \frac{\bar g^2}4 v^2 =
 \frac{m^2_{H_d} - m^2_{H_u} \tan^2 \beta}{\tan^2 \beta -1} \equiv M^2
\label{ext:1}
\end{equation}
\begin{equation}
\frac{\sin 2\beta}2 = 
\frac{\lambda L - \lambda A_\lambda x - B_\mu \mu'}
{m^2_{H_d} + m^2_{H_u} + 2 {\mu'}^2 + \lambda^2 v^2}
\label{ext:2}
\end{equation}
\begin{equation}
\mu' = \frac{\lambda \rho - \lambda^2 A_\lambda v_d v_u}
{ \lambda^2 v^2 +  m^2_S}, 
\label{ext:3}
\end{equation}
where $\tan \beta$ is defined as $\tan \beta = v_u/v_d$.
Note that the value of $\mu$ is redefined as $\mu' = \mu + \lambda x$.

Eqs. (\ref{ext:2}) and (\ref{ext:3})
lead to a cubic equation with respect to $v^2$, 
\begin{equation}
\label{cubic}
        (\lambda^2 v^2)^3 - 
         (\frac{2 \lambda L}{\sin 2 \beta} - m_{H_d}^2 - m_{H_u}^2)
         (\lambda^2 v^2)^2
   + 2 ( \lambda \rho )^2 =0,
\end{equation}
similarly to Eq. (\ref{cubic_eq}),
where we neglect $A_\lambda$, $B_\mu$ and $m_S$.\footnote{The parameter 
$A_\lambda$, $B_\mu$ and $m_S$ 
does not change the result as long as 
they are not comparable with the scale 10 TeV.}
The vacuum expectation value of the Higgs field
is
\begin{equation}
    v^2 \simeq \frac{2 L}{\lambda \sin 2\beta}. 
\end{equation}
Thus, the value for $v$ is larger than 10 TeV as a result.

%We attempt to modify the sliding singlet model
%to make it phenomenologically viable.
%We argue two possibilities to modify 
%the consequences (\ref{consequence}).

We cannot adopt the models where 
$F_X$ arises at the messenger level directly.
We, therefore, adopt the models where 
the $F$ component 
in the messenger sector arises radiatively.
It is possible that the Higgs field has a vacuum expectation 
value of weak scale if
the ratio $B = \rho/L$ is smaller than 100 GeV.
Suppose that the spurious
superfield Z in the supersymmetry breaking sector 
dominates the parameter $L$,
then we find the ratio $B$ to be $F_Z/Z$,
the lower bound 
for the expectation value of $Z$.
On the other hand, demanding the parameter $L$ 
to be less than the weak scale,
we find the upper bound for $Z$.
The parameter region to satisfy both boundaries exists.

%One of the possibilities is to make another spurious field $Z$
%for supersymmetry breaking.
%The largest vacuum expectation value of $F$ terms mainly gives 
%the parameter $\rho$, and similarly the largest
%value of $X F_X$ gives the parameter $L$.  
%We suppose that the field $Z$ gives the largest values
%for the components, $Z$ and $Z F_Z$;
%namely $Z$ dominates the parameter $\rho$ and $L$.
%The ratio $B = F_Z/Z$ for just the field $Z$ does not 
%necessarily dominate the supersymmetry breaking masses.
%The reason is that the
%supersymmetry breaking masses, which must be of the order of 
%$10^2$ GeV, are given by the largest ratio $B=F_X/X$ 
%of the fields coupling with messenger quarks directly.
%In other words, 
%the bound that the ratio $B$ has to be $10^4-10^5$ GeV 
%is not applied to the field $Z$ 
%in the case where
%the field $Z$ does not couple with messenger quarks directly or
%in the case where
%such fields exist that have larger ratio $B$ than $Z$ has. 
%Therefore, if it is satisfied that
%\begin{equation}
%    Z F_Z > X F_X, \quad F_Z > F_X \quad 
%   {\rm and} \quad \frac{F_Z}{Z} < 100 {\rm \ GeV},
%\end{equation}
%it is possible that Higgs has an vacuum expectation value
%less than 100 GeV.
%Such a situation is realized, 
%for example, in the models in which the 
%supersymmetry breaking sector and the messenger sector are separated.

One possibility to avoid the disappointing result of
 Eq. (\ref{consequence}) is to introduce an additional linear term
in the superpotential.
The linear term can arise dynamically
in $N_c = N_f$ Supersymmetric QCD, 
\begin{equation}
    W = S {\rm \, tr} M + \bar \mu (\det M - B \bar B - \Lambda^{2N_c}),
\end{equation}
where $\bar \mu$ is a Lagrange multiplier.
Integrating out the meson field $M$, we obtain the linear term
of $S$, 
\begin{equation}
    W = S \Lambda^2.
\end{equation}
When
\begin{equation}
    \sqrt{L} < \Lambda < 100 {\rm \ GeV},
\end{equation}
it is possible that Higgs has an vacuum expectation value
less than 100 GeV.
In this case, the gravitino mass has to be less than 10 keV.

In any case, however, 
it is difficult to construct a phenomenologically viable model.
The main reason is that the gauge mediation models produce
large $\mu$ term at low energy.

%
%
%Eq. (\ref{ext:1}) states the necessity for  
%the cancellation between ${\mu'}^2$ and
%$m_{H_u}^2$ as phenomenological constraint.
%

The most stringent phenomenological constraint 
on the gauge mediation models is derived from
the lower bound of the right-handed scalar electron mass, that is,
$80$ GeV \cite{80gev}.
This constraint gives the lower bound of the parameter $N_m \Lambda_{mes}^2$
because the supersymmetry breaking 
scalar mass squared is proportional to 
$N_m (\alpha_i/4\pi)^2 \Lambda_{mes}^2$ at the messenger scale \cite{GM},
where $N_m$ is the number of messenger quarks or leptons.
If the messenger scale is not so large
({\it e.g.} smaller than $10^6$ GeV or so), this
parameter $N_m \Lambda_{mes}^2$ has to be greater than about 
$(40 {\rm TeV})^2$.\footnote{The RGE correction is dependent 
on the messenger scale.
The right-handed scalar electron mass grows up as raising the messenger scale.
However, the messenger scale can not be larger than $10^{10}$ GeV 
in our scheme because
the parameters $L$ and $\rho$ must not be larger than the weak scale.
Such a low messenger scale does not change the condition that 
the parameter $N_m \Lambda_{mes}^2$ should be 
larger than about $(40 {\rm TeV})^2$.}
%If one imposes that the right-handed scalar electron mass is larger
%than 80 GeV, the parameter $N_m \Lambda_{mes}^2$ has to be 
%larger than $(39 {\rm TeV})^2$.\footnote{The RGE correction 
%is dependent on the messenger scale.
As such, $N_m \Lambda_{mes}^2$ makes soft supersymmetry breaking
 right-handed scalar top mass squared
 larger than about $(430 {\rm GeV})^2$.
%\footnote{This "scalar top mass" does not mean
%the eigenvalues of the scalar top mass matrix, 
%but only the parameters which 
% appear in the RGE.}
%\footnote{In this statement, only diagonal 
%elements of the scalar top mass matrix are considered.
%Though eigenstates of this mass matrix can be lower 
%than $430$ GeV, the
%result of RGE that $m_{H_u}^2$ becomes lower than 
%$-(200 {rm \ GeV})^2$ does
%not change.}
Furthermore, this large soft supersymmetry breaking scalar 
top mass squared drives $m_{H_u}^2$ 
lower than typically minus $(200 {\rm \ GeV})^2$ through 
renormalization group equations\cite{GoFrMu, finetune}.
In order to yield Z boson mass, ${\bar g} v/\sqrt{2}$, of the magnitude 
91 GeV, 
the $\mu'$ parameter should be larger than 190 GeV 
from Eq. (\ref{ext:1}).
Even in the larger messenger scale,
the lower bound of $\mu'$ parameter 
is conservatively about 160 GeV \cite{GoFrMu}.
%In the MSSM, at least $M_Z^2/2\mu'^2 = 16\%$ cancellation 
%between $\mu'^2$ and $m_{H_u}^2$ is 
%needed to reproduce the observed $M_Z$.
%A more accurate cancellation is required for most of the parameter 
%space \cite{GoFrMu}.

%In the models in which the $\mu$ parameter is replaced by 
%a singlet field, even this fine-tuning becomes impossible.
%The extremization solution that satisfies 
%$M_Z^2 \ll {\mu'}^2$ is not on the minimum point but on the saddle point.
This large $\mu'$ makes the extremization solution unstable because it
is on a saddle point.
Neglecting $A_\lambda$ and $B_\mu$, 
we obtain the scalar mass squared matrix ${\cal M}^2$, 
\begin{eqnarray}
{\cal M}^2 &\!\!\!=& \!\!\! \frac{1}{2} 
                \frac{\partial^2
                V}{\partial v_i \partial v_j} \nonumber \\
           &\!\!\!=&\!\!\! \frac{1}{2} \left(\begin{array}{ccc}
                2 \lambda^2 v^2 + 2 m_S^2 &
                4 \lambda \mu' v_d &
                4 \lambda \mu' v_u \\
                4 \lambda \mu' v_d &
                \bar{g}^2 v_d^2 + 2 \lambda L \tan \beta &
                (4\lambda^2 - \bar{g}^2)v_d v_u - 2 \lambda L \\
                4 \lambda \mu' v_u &
                (4\lambda^2 - \bar{g}^2)v_d v_u - 2 \lambda L &
                \bar{g}^2 v_u^2 + 2 \lambda L \cot \beta
                \end{array}\right), \nonumber \\
\end{eqnarray}
where we define $v_1=x$, $v_2=v_d$ and $v_3=v_u$.
Determinant of this matrix is
%\newpage
\begin{eqnarray}
\det{\cal M}^2 &\!\!\!=&\!\!\! 
               4 v^2 
                      \left\{ 2 \left[ (m_{H_u}^2 + m_{H_d}^2 + 2 {\mu'}^2)
                      \left( \bar{g}^2 \cos^2 2 \beta  + 
                      2 \lambda^2 \sin^2 2\beta \right) 
                      + \bar{g}^2 \lambda^2 v^2\right] m_S^2 \right. 
                      \nonumber \\
              &   & \ \ \ 
                      -\lambda^2 \left[ 2(2{\mu'}^2 + m_{H_d}^2 + m_{H_u}^2
                      + M_Z^2)(4 {\mu'}^2 - \lambda^2 v^2)
                      \right. \nonumber \\
              &  &  \ \ \ \ \ \ \ \ \:
                      \left.\left. +(2\lambda^2 - \bar{g}^2)v^2
                      (6{\mu'}^2 + m_{H_d}^2 + m_{H_u}^2)\cos^2 2\beta
                      \right]\right\}.
\end{eqnarray}
One can easily find that the expression inside the second square 
brackets is positive 
provided $\bar{g}^2 v^2, \lambda^2 v^2 \ll {\mu'}^2$ and $|m_{H_u}^2|$.
Actually, this expression is positive even when $\mu' \sim \lambda v$ 
and $\mu' \sim \bar{g} v$.
Thus, the phenomenologically viable solution
can not lie on the minimum point
unless 
the supersymmetry breaking mass squared of the singlet, 
$m_S^2$, is larger than at least the order of 
${\mu'}^2$.
However, the gauge mediation model does not make such a large breaking mass
of the gauge singlet.
We need some extra mechanism for making such a large $m_S^2$.

Though we find that such a mechanism yields a phenomenologically
 viable solution, 
 we need fine-tuning between $m_S^2$ and $\rho$.
In the case that $m_S^2$ is large, from Eqs. (\ref{ext:1}) and 
(\ref{ext:3}), we study the cubic equation 
for $v^2$ to find what fine-tuned parameters
 are needed.
Neglecting $A_{\lambda}$ and $B_\mu$, we obtain 
\begin{equation}
X^3 + \left( \frac{\bar{g}^2 m_S^2}{2 \lambda^2 M^2}-1 \right) X^2 +
\frac{\bar{g}^2 m_S^2}{4 \lambda^2 M^2}\left( \frac{\bar{g}^2 m_S^2}
{4 \lambda^2 M^2} -2 \right) X + \left( \frac{\bar{g}^2 \rho}{4 \lambda M^3}
\right)^2 - \left(\frac{\bar{g}^2 m_S^2}{4 \lambda^2 M^2}\right)^2 = 0,
\end{equation}
where we define a dimensionless variable $X \equiv \bar{g}^2 v^2/(4 M^2)$.
The phenomenologically viable solution is then
\begin{equation}
X <  \frac{(91{\rm \ GeV})^2}
{2 \times (200\rm{\ GeV})^2} \sim 0.1.
\end{equation}
We need still a fine-tuned parameter such as 
$m_S^2 \sim \lambda \rho/M$ for such a small $X$ solution, 
even if we find an extra mechanism for making large $m_S^2$.

%
%%%%%%%%%  Discussion  %%%%%%%%%%%%%%%%%%%%%%%%%%%%%%%%%%%%%%
%
\vspace{7mm}
\pagebreak[3]
\addtocounter{section}{1}
\setcounter{equation}{0}
\setcounter{subsection}{0}
\setcounter{footnote}{0}
\begin{center}

{\large {\bf \thesection. Conclusion
}}
\end{center}
\nopagebreak
\medskip
\nopagebreak
\hspace{3mm}

In this paper, we analyze the sliding singlet mechanism
proposed to solve the fine tuning problem associated with 
the triplet-doublet splitting problem
in the gauge mediation scenario. 
The supersymmetry breaking effects induce 
the singlet linear terms both in the superpotential 
and in the soft supersymmetry breaking interactions. 
We show that these terms change 
the potential minimum drastically, which 
cause the weak scale to be of $O(10)$ TeV in the direct
 gauge mediation model.
Even in the other models, we should choose rather extreme parameters,
 $F_Z/Z < O(10^2)$ GeV in the supersymmetry breaking sector,
 and $F_X/X > 10^4$ GeV in the messenger sector.
We analyzed the minimal effects derived from 
the supersymmetry breaking in the messenger sector, 
which always exist in all messenger models. 
Our analysis was applied to all gauge mediation models.
From this analysis, it was found that the sliding singlet mechanism 
in the gauge mediation scenario 
is extremely constrained
unless the model is extended. 
%{}For example, we consider the situation that 
%there is another singlet $Z$
%in the dynamical supersymmetry breaking sector
%which satisfies $Z F_Z 
%> X F_X$, $F_Z 
%> F_X$ and $F_Z / Z 
%< O(10^2)$ GeV.
We can avoid the difficulty by, for example, 
introducing the additional gauge group
whose non-perturbative effects induce 
the linear term in the superpotential but 
do not induce the soft breaking linear term.
However, even if we extend the model, 
we also need one more additional mechanism 
that induces such a sliding singlet soft breaking mass 
as $m_S^2 \sim O(\mu^2)$.
Moreover, this $m_S^2$ must be fine-tuned as 
$m_S^2 \sim \lambda \rho /M$ 
to obtain the correct weak scale. 
This corresponds to fine-tuning between
the $\mu$ term and soft masses of Higgs scalars. 
However, this fine-tuning is not special to our model, 
but appears even in the 
minimal supersymmetric standard model \cite{GoFrMu}.

%
%
%%%%%%%%%%%%%%%%%%%%%%%% note added in proof%%%%%%%%%%%%%%%%%%%%%%%%%%%%
%
%
\vspace{7mm}
\pagebreak[3]
\addtocounter{section}{1}
\setcounter{equation}{0}
\setcounter{subsection}{0}
\setcounter{footnote}{0}
\begin{center}

{\large {\bf Note added in proof
}}
\end{center}
\nopagebreak
\medskip
\nopagebreak
\hspace{3mm}

The careful reader may point out the possibility that a new minimum appears
under the messenger scale through the quantum effect.
Renormalization group equation changes scalar potential at the point
$\lambda S = 3 \lambda' \sigma$, on which the masses of Higgs doublets
are zero.
Correct minimum of minimal supersymmetric standard models (MSSM)
appears at least as the local minimum.
On the other hand, true minimum of the scalar potential
 above the messenger scale 
is on the point $S=\rho/m_{3/2}^2$ and $v=0$.
Thus, the condition for which the MSSM minimum becomes a true minimum is 
$$V(\lambda S = 3 \lambda' \sigma , v=0) - V(\lambda S = \rho/m_{3/2}^2,
 v=0) < (10^2 {\rm GeV})^4, \eqno(A.1)$$
where $V$ is the same $V$ as in Eq. (\ref{scalarpot}).
This condition demands the gravitino mass to be less than $10^{-3}$ eV.
Such a small gravitino mass requires the 
SUSY breaking order parameter $\sqrt{F} < 10^3$  GeV.
This is incompatible with gauge mediation models.

%
%
%%%%%%%  Acknowledgement  %%%%%%%%%%%%%%%%%%%%%%%%%%%%%%%%%%%
%
\pagebreak[3]
\addtocounter{section}{1}
\setcounter{equation}{0}
\setcounter{subsection}{0}
\setcounter{footnote}{0}
\begin{center}
{\large {\bf  Acknowledgement
}}
\end{center}
\nopagebreak
\medskip
\nopagebreak
\hspace{3mm}
This work was partially supported by 
JSPS Research Fellowships for Young Scientists (Y.C. and Y.M.).
%
%
%%%%%%%  References  %%%%%%%%%%%%%%%%%%%%%%%%%%%%%%%%%%%%%%%
\vspace{5mm}
%\newpage
%
\newcommand{\NP}[1]{{\it Nucl.\ Phys.\ }{\bf #1}}
\newcommand{\PL}[1]{{\it Phys.\ Lett.\ }{\bf #1}}
\newcommand{\CMP}[1]{{\it Commun.\ Math.\ Phys.\ }{\bf #1}}
\newcommand{\MPL}[1]{{\it Mod.\ Phys.\ Lett.\ }{\bf #1}}
\newcommand{\IJMP}[1]{{\it Int.\ J. Mod.\ Phys.\ }{\bf #1}}
\newcommand{\PRP}[1]{{\it Phys.\ Rep.\ }{\bf #1}}
\newcommand{\PR}[1]{{\it Phys.\ Rev.\ }{\bf #1}}
\newcommand{\PRL}[1]{{\it Phys.\ Rev.\ Lett.\ }{\bf #1}}
\newcommand{\PTP}[1]{{\it Prog.\ Theor.\ Phys.\ }{\bf #1}}
\newcommand{\PTPS}[1]{{\it Prog.\ Theor.\ Phys.\ Suppl.\ }{\bf #1}}
\newcommand{\AP}[1]{{\it Ann.\ Phys.\ }{\bf #1}}
\newcommand{\ZP}[1]{{\it Zeit.\ f.\ Phys.\ }{\bf #1}}

\begin{thebibliography}{100}
%
%
\bibitem{sliding} E. Witten, \PL{B105} (1981) 267. \\
D. V. Nanopoulos and K. Tamvakis, \PL{B113} (1982) 151. \\
S. Dimopoulos and H. Georgi, \PL{B117} (1982) 287. \\
L. Ibanez and G. Ross, \PL{B110} (1982) 215.
%
\bibitem{Neme} D. Nemeschansky, \NP{B234} (1984) 379.
%
\bibitem{missing} H. Georgi, \PL{B108} (1982) 283. \\
A. Masiero, D. V. Nanopoulos, K. Tamvakis and T. Yanagida, 
\PL{B115} (1982) 380, \\
B. Grinstein, \NP{B206} (1982) 387. 
%
\bibitem{DW} S. Dimopoulos and F. Wilczek, 
Preprint NSF-ITP-82-07 (1982). \\
K. S. Babu and S. M. Barr, \PR{D48} (1993) 5354.  
%
\bibitem{GIFT} K. Inoue, A. Kakuto and T. Takano, 
\PTP{75} (1986) 664. \\
A. Anselm and A. Johansen, \PL{B200} (1988) 331. \\
A. Anselm, {\it Sov. Phys. JETP} {\bf 67} (1988) 663. \\
Z. G. Berezhiani and G. Dvali, {\it Sov. Phys. Lebedev Inst. Reports} 
{\bf 5} (1989) 55.
%
\bibitem{radiative}
K. Inoue, A. Kakuto, H. Komatsu and S. Takeshita,
\PTP{68} (1982) 927; \PTP{71} (1984) 413. \\
L. E. Ib\'a\~nez, \PL{B118} (1982) 73;
\NP{B218} (1983) 514. \\
L. Alvarez-Gaum\'e, J. Polchinski and M. Wise, \NP{B221} (1983) 495.
%
\bibitem{PS} J. Polchinski and L. Susskind, \PR{D26} (1982) 3661. \\
M. Dine, ``{\it Lattice Gauge Theories, Supersymmetry, and 
Grand Unification,} Proc. of the 6th Johns Hopkins Workshop, 
Florence, Italy, 1982 (Johns Hopkins Press, Baltimore, 1982). \\
H. P. Nilles, M. Srednicki and D. Wyler, \PL{B124} (1982) 337. \\
A. B. Lahanas, \PL{B124} (1982) 341. 
%
\bibitem{Sen} A. Sen, \PL{B148} (1984) 65. \\
S. M. Barr, \PR{D57} (1998) 190.                 
%
\bibitem{GM} 
M. Dine and W. Fischler, \PL{B110} (1982) 227.\\
C. R. Nappi and B. A. Ovrut, \PL{B113} (1982) 175.\\
L. Alvarez-Gaum\'{e}, M. Claudson and M. B. Wise, \NP{B207} (1982) 96.\\
M. Dine and A. E. Nelson, \PR{D48} (1993) 1277. \\
G. F. Giudice and R. Rattazzi, hep-ph/9801271. 
%
\bibitem{DNS}
M. Dine, A. Nelson, Y. Nir and Y. Shirman, \PR{D53} (1996) 2658. 
%
\bibitem{CiPo} P. Ciafaloni and A. Pomarol, \PL{B404} (1997) 83.
%
\bibitem{Nilles} 
J. Bagger and E. Poppitz, \PRL{71} (1993) 2380. \\
V. Jain, \PL{B351} (1995) 481. \\
J. Bagger, E. Poppitz and L. Randall, \NP{B455} (1995) 59. \\
H. P. Nilles and N. Polonsky, \PL{B412} (1997) 69. \\
C. Kolda, S. Pokorski and N. Polonsky, \PRL{80} (1998) 5263.
%
\bibitem{80gev} Talk presented by Glen Cowan (ALEPH collaboration) at the 
special CERN particle physics seminar on physics results from the LEP
run at 172 GeV, 25 February, 1997.\\
G. F. Giudice, M. L. Mangano, G. Ridolfi and R. Ruckel ({\it convenors}),
{\it Searches for New Physics}, hep-ph/9602207.
%
\bibitem{GoFrMu} A. de Gouv\^{e}a, A. Friedland and H. Murayama,
\PR{D57} (1998) 5676. 
%
\bibitem{finetune} K. Agashe and M. Graesser, \NP{B507} (1997) 3. \\
P. Ciafaloni and A. Strumia, \NP{B494} (1997) 41.
%
%
%\bibitem{cosmo} T. Moroi, H. Murayama and M. Yamaguchi, \PL{B303} (1993) 289.
%
\end{thebibliography}
\end{document}